\documentclass[twocolumn,showpacs,aps]{revtex4}
\usepackage{graphicx}
\usepackage{xcolor}
\usepackage{subfigure}
\usepackage{amssymb}
\usepackage{amsmath}

\begin{document}

\title{Scaling laws for near barrier Coulomb and Nuclear Breakup }
\author{M.S. Hussein }
\affiliation{Instituto de Estudos Avan\c{c}ados, Universidade de S\~{a}o Paulo C. P.
72012, 05508-970 S\~{a}o Paulo-SP, Brazil, and Instituto de F\'{\i}sica,
Universidade de S\~{a}o Paulo, C. P. 66318, 05314-970 S\~{a}o Paulo,-SP}
\author{P. R. S. Gomes, J. Lubian, D. R.\ Otomar}
\affiliation{Instituto de F\'{\i}sica, Universidade Federal Fluminense, Av. Litoranea
s/n, Gragoat\'{a}, Niter\'{o}i, R.J., 24210-340, Brazil}
\author{ L. F. Canto}
\affiliation{Instituto de F\'{\i}sica, Universidade Federal do Rio de Janeiro, CP 68528,
Rio de Janeiro, Brazil \ }
\date{\today }

\begin{abstract}
We investigate the nuclear and the Coulomb contributions to the breakup
cross sections of $^6$Li in collisions with targets in different mass
ranges. Comparing cross sections for different targets at collision energies
corresponding to the same $E/V_{\mathrm{\scriptscriptstyle  B}}$, we obtain
interesting scaling laws. First, we derive an approximate linear expression
for the nuclear breakup cross section as a function of $A_{\mathrm{%
\scriptscriptstyle  T}}^{1/3}$. We then confirm the validity of this
expression performing CDCC calculations. Scaling laws for the Coulomb
breakup cross section are also investigated. In this case, our CDCC
calculations indicate that this cross section has a linear dependence on the
atomic number of the target. This behavior is explained by qualitative
arguments. Our findings, which are consistent with previously obtained
results for higher energies, are important when planning for experiments
involving exotic weakly bound nuclei.
\end{abstract}

\maketitle

\bigskip

\section{Introduction}

The effect of the coupling of the breakup channel on the complete fusion of
weakly bound, and especially, halo nuclei, is a subject that has been
extensively investigated in the last years, both experimentally and
theoretically\cite{Canto06}. There are strong signatures that this coupling
hinders the fusion cross section at energies above the barrier, and enhances
the tunneling-dominated fusion below the barrier\cite{Canto09JPG,
Canto09NPA, Gomes11PLB, Gomes12JPG}. Owing to the importance of the breakup
channel at low energies, close to the Coulomb barrier, it is very
interesting to investigate the dependence of the relative importance of its
Coulomb and nuclear components, $\sigma _{\mathrm{\scriptscriptstyle  C}}^{%
\mathrm{bu}}$ $\ $ and $\sigma _{\mathrm{\scriptscriptstyle  N}}^{\mathrm{bu}%
} $, respectively, on the mass number or charge of the target nucleus. There
are results showing that the above mentioned effects of breakup coupling on
fusion cross sections are less important for light systems than for heavy
ones~\cite{Gomes09}.

However, one cannot calculate the total breakup cross section $\sigma ^{%
\mathrm{bu}}$ as $\sigma ^{\mathrm{bu}}=\sigma _{\mathrm{\scriptscriptstyle  %
C}}^{\mathrm{bu}}+\sigma _{\mathrm{\scriptscriptstyle  N}}^{\mathrm{bu}}$
since the interference of these two components may be very strong, as
demonstrated in \cite{hussein06, otomar13,nunes98}. From $\sigma _{\mathrm{%
\scriptscriptstyle  C}}^{\mathrm{bu}}$ one can extract information about the
collective response of halo and other weakly bound nuclei, such as the
dipole response. By considering a fixed weakly bound projectile, one might
expect that $\sigma _{\mathrm{\scriptscriptstyle  C}}^{\mathrm{bu}}$ should
increase with the charge of the target. For heavy targets, the Coulomb
breakup should predominate over the nuclear breakup. On the other hand, the
dependence of $\sigma _{\mathrm{\scriptscriptstyle  N}}^{\mathrm{bu}}$\ with
the target nucleus characteristics is more difficult to predict. Two
reported works \cite{hussein06, acquadro81} deal with the study of the
nuclear breakup as a function of the target mass at high energies, around
bombarding energies of some tens of MeV/n. According to those works, the
nuclear breakup cross section behaves at a given value of the bombarding
energy, $E_{\mathrm{lab}}$, as,

\begin{equation}
\sigma^{\mathrm{bu}}_{\mathrm{\scriptscriptstyle  N}} = P_1 + P_2 \, A_{%
\mathrm{\scriptscriptstyle  T}}^{1/3}.
\end{equation}
Above, $P_{1}$ and $P_{2}$ are functions of the bombarding energy and the
structure of the projectile. This formula, well substantiated by extensive
continuum discretized coupled channel calculations (CDCC) performed in Ref. 
\cite{hussein06}, was used to estimate the nuclear breakup cross section for
a heavy target, using the experimental value of the cross section for
collisions of the same projectile with a light target. In this case, the
Coulomb breakup is much smaller and to a first approximation it can be
neglected. Frequently~\cite{naka99, naka94,aum99,Leisten01}, the nuclear
contribution to the breakup cross section is estimated in this way and the
Coulomb contribution is obtained subtracting this contribution from the
experimental total breakup cross section. The Coulomb breakup cross section
obtained in this way can be compared with the one given by expression~\cite%
{bertu88, bertu01, naka99} , 
\begin{equation}
\sigma _{\mathrm{\scriptscriptstyle  C}}^{\mathrm{bu}}=\frac{16 \pi ^{3}}{9}%
\ \alpha \int dE_{x}\ n_{E1}(E_{x})\ \frac{dB(E1)}{e^{2}\,dE_{x}},
\end{equation}
where $\alpha$ is the fine structure constant, $n_{E1}$ is the number of
virtual photons, and $dB(E1)/e^{2}\, dE_{x}$ is the dipole response. Since $%
n_{E1}$ is known, the above expression for $ \sigma _{\mathrm{%
\scriptscriptstyle  C}}^{\mathrm{bu}}$ is used to test models of the dipole
response. However, assuming that the Coulomb breakup cross section can be
given by the difference between the total and the nuclear breakup cross
section may be very wrong. This procedure neglects the interference between
the Coulomb and the nuclear breakup amplitudes, which may be quite important~%
\cite{hussein06}.

\bigskip

In Ref. \cite{hussein06}, Hussein \textit{el al.} performed CDCC
calculations for the nuclear breakup of $^{8}$B, $^{11}$Be, and $^{7}$Be
projectiles in collisions with several targets. The cross sections $\sigma_{%
\mathrm{\scriptscriptstyle  N}}^{\mathrm{bu}}$ for different collision
energies were plotted against the target mass. They concluded that the
non-halo projectile $^{7}$Be seemed to obey the scaling law quite well.
However, the halo nuclei required values and signs of $P_{1}$ and $P_{2}$
which were not consistent with the simple scaling law obtained at higher
energies.

Very recently we reported \cite{otomar13}\ results of a study on this
subject at lower energies, around the Coulomb barrier. We investigated the
breakup process evaluating separate contributions from the Coulomb and from
the nuclear fields, as well as the Coulomb-nuclear interference, through
CDCC\ calculations. We performed calculations for collisions of the $^{6}$Li
projectile with $^{59}$Co, $^{144}$Sm and $^{208}$Pb, at three energies very
close to the Coulomb barrier ($E/V_{\mathrm{\scriptscriptstyle  B}}$ = 0.84,
1.00 and 1.07, where $V_{\mathrm{\scriptscriptstyle  B}}$ is the Coulomb
barrier). The choice of these systems was based on the availability of
elastic scattering data in the literature. In this way, we were able to
check the reliability of our calculations, which do not contain any
adjustable parameter, through comparisons with the scattering data. The
results showed a linear dependence of the Coulomb breakup cross section, $%
\sigma _{\mathrm{\scriptscriptstyle  C}}^{\mathrm{bu}}$, with the charge of
the target, for the same $E/V_{\mathrm{\scriptscriptstyle  B}}$\ values. A
linear dependence of the nuclear breakup cross section, $\sigma _{\mathrm{%
\scriptscriptstyle  N}}^{\mathrm{bu}}$, was found as a function of $A_{%
\mathrm{\scriptscriptstyle  T}}^{1/3}$, similar to what was found at high
energies \cite{hussein06}, but for the same values of $E/V_{\mathrm{%
\scriptscriptstyle  B}}$, instead of $E_{\mathrm{lab}}$. Furthermore, we
have shown a strong interference between the two breakup components, in such
way that $\sigma ^{\mathrm{bu}}$\ is smaller than $\sigma _{\mathrm{%
\scriptscriptstyle  C}}^{\mathrm{bu}}+\sigma _{\mathrm{\scriptscriptstyle  N}%
}^{\mathrm{bu}}$\ for all systems and all energies investigated and, for
sub-barrier energies $\sigma _{\mathrm{\scriptscriptstyle  C}}^{\mathrm{bu}}$%
\ is even larger than $\sigma ^{\mathrm{bu}}$.

However, in ref. \cite{otomar13} we did not derive theoretically the
dependence of $\sigma _{\mathrm{\scriptscriptstyle  N}}^{\mathrm{bu}}$\ with
the target mass for the same $E/V_{\mathrm{\scriptscriptstyle  B}}$\ values,
and the calculations were restricted to energies very close to $V_{\mathrm{%
\scriptscriptstyle  B}}$. So, in this Brief Report, as a complement to ref. 
\cite{otomar13}, we show the theoretical derivation of $\sigma _{\mathrm{%
\scriptscriptstyle  N}}^{\mathrm{bu}}$\ as a function of the target mass and
we compare those results with similar CDCC calculations. Here, we consider
collision energies in a wider range, from $E/V_{\mathrm{\scriptscriptstyle  B%
}}$ = 0.84 to 3.0, and we also fill the gap existing between $^{59}$Co and $%
^{144}$Sm target nuclei, including in our investigation the $^{6}$Li + $%
^{120}$Sn system.

\medskip

In the following we derive the low energy version of Eq. (1) for the nuclear
breakup. We start with the Wong formula for the fusion cross section taken
to be the total nuclear reaction cross section, without breakup \cite{Wong},

\begin{equation}
\sigma_{\mathrm{\scriptscriptstyle  F}} = \frac{\Gamma}{ E}\ \pi R^2\ \ln{%
\left[1 + \exp{\left( \frac{\hbar^{2}\Lambda_{\mathrm{\scriptscriptstyle  C}%
}^2}{2\mu R^2\ \Gamma}\right)}\right]},
\end{equation}
where $\Gamma = \hbar \omega/2\pi$ is an energy width related to the
curvature of the Coulomb barrier ($\hbar\omega)$ and $R$ barrier radius. The
variable $\Lambda_{\mathrm{\scriptscriptstyle  C}}$ is the critical angular
momentum associated with fusion.

The reaction cross section including nuclear breakup but not Coulomb breakup
(this contribution is of a long range nature and can not be described using
the Wong formula), can be written as,

\begin{equation}
\sigma_{\mathrm{\scriptscriptstyle  R}} = \frac{\Gamma}{E}\ \pi R^2\ \ln{%
\left[1 + \exp{\left( \frac{\hbar^2(\Lambda_{\mathrm{\scriptscriptstyle  C}}
+ \Delta)^2}{2\mu R^2\ \Gamma}\right)}\right]}.
\end{equation}

\noindent The nuclear breakup cross section is taken to be the difference 
\cite{Glas}, 
\begin{equation}
\sigma _{\mathrm{\scriptscriptstyle N}}^{\mathrm{bu}}=\sigma _{\mathrm{%
\scriptscriptstyle R}}-\sigma _{\mathrm{\scriptscriptstyle F}}=\frac{\Gamma 
}{E}\ \pi R^{2}\ \ln {\ \left[ \frac{1+\exp {\left( \frac{\hbar ^{2}(\Lambda
_{\mathrm{\scriptscriptstyle C}}+\Delta )^{2}}{2\mu R^{2}\ \Gamma }\right) }%
}{1+\exp {\left( \frac{\hbar ^{2}\Lambda _{\mathrm{\scriptscriptstyle C}}^{2}%
}{2\mu R^{2}\ \Gamma }\right) }}\right] }.
\end{equation}%
This expression can be simplified by expanding to lowest order in $\Delta
/\Lambda _{\mathrm{\scriptscriptstyle C}}$, to give,

\begin{multline}
\sigma^{\mathrm{bu}}_{\mathrm{\scriptscriptstyle  N}} = 2\,\frac{\pi}{k^2}\,
\Lambda_c \Delta = \\
2\pi a\, \left[1 - \frac{V_{\mathrm{\scriptscriptstyle  B}}}{E}\right] (R_{%
\mathrm{\scriptscriptstyle  P}} + R_{\mathrm{\scriptscriptstyle  T}}) = P_1
+ P_2 A_{\mathrm{\scriptscriptstyle  T}}^{1/3},  \label{P1P2}
\end{multline}
\noindent where we have used $\Lambda_{\mathrm{\scriptscriptstyle  C}}^2 =(
2\mu/\hbar^2)( R_{\mathrm{\scriptscriptstyle  P}} + R_{\mathrm{%
\scriptscriptstyle  T}})^2 [E - V_{\mathrm{\scriptscriptstyle  B}}] = k^2 [1
- V_{\mathrm{\scriptscriptstyle  B}}/E](R_{\mathrm{\scriptscriptstyle  P}} +
R_{\mathrm{\scriptscriptstyle  T}})^2$, and $\Delta^2 = k^2[1 - V_{\mathrm{%
\scriptscriptstyle  B}}/E]\ a^2 $, with $a$ being the diffuseness of the
nuclear surface. Clearly $P_1 = 2\pi \,a\, (1 - V_{\mathrm{%
\scriptscriptstyle  B}}/E)R_{\mathrm{\scriptscriptstyle  P}}$ and $P_2 =
2\pi a\, r_0 (1 - V_{\mathrm{\scriptscriptstyle  B}}/E)$.

\medskip

Eq.~(\ref{P1P2}) clearly shows that for a fixed $E/V_{\mathrm{%
\scriptscriptstyle  B}}$ and for a given projectile, the cross section
scales linearly with the radius of the target, as it was verified in ref.~%
\cite{otomar13}. The above formula for the nuclear breakup cross section
represents the scaling at low energies and reduces to the one discussed in 
\cite{hussein06} at higher energies. Of course the factor $(1-V_{\mathrm{%
\scriptscriptstyle  B}}/E)$ is only meaningful at above-barrier energies. As
the energy is lowered below the barrier, tunneling takes over and one must
rely on a different approximation. Once we have derived this scaling law,
the next step is to confirm its validity by calculating the nuclear breakup
cross sections through CDCC\ calculations. The CDCC method \cite{KYI86,AIK87}
is the most suitable approach to deal with the breakup process, which feeds
states in the continuum, whose wave functions are grouped in bins or wave
packets that can be treated similarly to the usual bound inelastic states,
since they are described by square-integrable wave functions. In the present
work we extend the calculations already presented in ref. \cite{otomar13},
and so we will not repeat the details here, since they can be found in that
reference and in Refs.~\cite{KYI86,AIK87,Otomar10}. Only the main aspects of
the method and some specific details of the calculations will be mentioned
in the following. It is assumed that $^{6}$Li projectiles breakup into a
deuteron and an alpha particle, and so it is used the cluster model in which 
$^{6}$Li is described as a bound state of the $d+\alpha $ system and the
breakup channel is represented by the continuum states of this system, as it
was successfully done in previous works~\cite{SYK82,KRu96}. The calculations
were performed using the computer code FRESCO~\cite{Tho88}. In the cluster
model, the projectile-nucleus interaction is written as

\begin{equation}
V(\mathbf{R},\mathbf{r},\xi )=V_{\alpha -{\mathrm{\scriptscriptstyle T}}}(%
\mathbf{R},\mathbf{r},\xi )+V_{d-{\mathrm{\scriptscriptstyle T}}}(\mathbf{R},%
\mathbf{r},\xi ),  \label{potentials}
\end{equation}
where \textbf{R} is the vector joining the projectile's and target's
centers, \textbf{r} is the relative vector between the two clusters ($d$ and 
$\alpha $), and $\xi$ stands for any other intrinsic coordinate describing
the projectile-target system. The continuum states of $^{6}$Li are
discretised as in Refs.~\cite{DTB03,OLG09,Otomar10}. The interaction between
the $d$ and the $\alpha $ clusters within $^{6}$Li is given by a Woods-Saxon
potential, with the same parameters as in Refs.~\cite{DTB03,OLG09,Otomar10}.
The real parts of the $V_{\alpha -{\mathrm{\scriptscriptstyle T}}}(\mathbf{R}%
,\mathbf{r},\xi )$ and $V_{d-{\mathrm{\scriptscriptstyle T}}}(\mathbf{R},%
\mathbf{r},\xi )$ interactions are given by the double-folding S\~{a}o Paulo
potential~\cite{Chamon}. We have assumed that the mass densities of the $d$
and $\alpha $ clusters, required for the double-folding calculation, can be
approximated by the charge densities multiplied by two, whereas the mass
densities of the targets were taken from the systematic study of Ref.~\cite%
{Chamon}. The imaginary parts of $V_{\alpha -{\mathrm{\scriptscriptstyle T}}%
}(\mathbf{R},\mathbf{r},\xi )$ and $V_{d-{\mathrm{\scriptscriptstyle T}}}(%
\mathbf{R},\mathbf{r},\xi )$ were chosen as to represent short-range fusion
absorption, corresponding to assume ingoing wave boundary conditions. The
CDCC calculations include also inelastic channels, corresponding to
collective excitations of the targets. The channels selected for the three
targets reported in ref. \cite{otomar13} were already descibed in that
paper. For the $^{120}$Sn nucleus, the excitation included was the
one-phonon quadrupole (2$^{+}$, E$^{\ast }$ = 1.1714 MeV) first order
vibration. The values of the deformation parameters were obtained from Ref.~%
\cite{Ram01} and \cite{Kib02} for the quadrupole and octupole deformations,
respectively.

\begin{figure}[ptb]
\begin{center}
\includegraphics*[height=7cm]{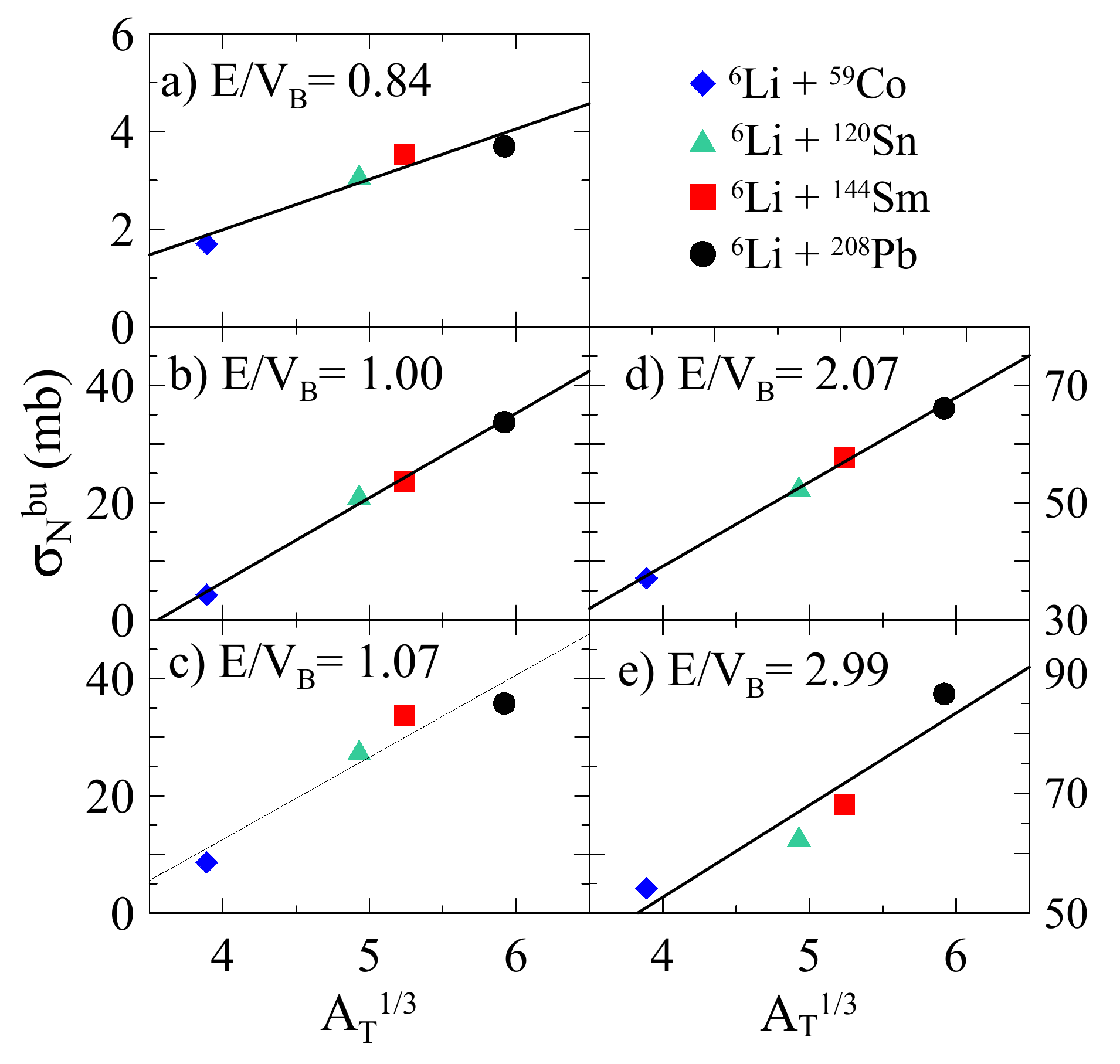}
\end{center}
\caption{(colour on line) Nuclear breakup cross sections plotted as
functions of $A^{1/3}_{\mathrm{\scriptscriptstyle  T}}$, for the energies $%
E/V_{\mathrm{\scriptscriptstyle  B}}=0.84$ (panel a), 1.00 (panel b), 1.07
(panel c), 2.07 (panel d) and 2.99 (panel e). The straight lines are linear
fits to the data.}
\label{fig1}
\end{figure}
\bigskip Figure~\ref{fig1} shows the nuclear breakup cross sections obtained
with our CDCC calculations, as functions of $A_{\mathrm{\scriptscriptstyle  T%
}}^{1/3}$. The systems are the ones mentioned above and we consider 5
different energies in panels a), b), c), d) and e). The notation is
explained in the figure. The straight lines are linear fits to the points.
We see that the results are very well fitted by the lines in all cases, in
agreement with the expression of Eq.~(\ref{P1P2}). One observes that
although this equation was derived for energies above the barrier, it seems
to remain valid even slightly below the barrier.

\begin{figure}[ptb]
\begin{center}
\includegraphics*[height=7cm]{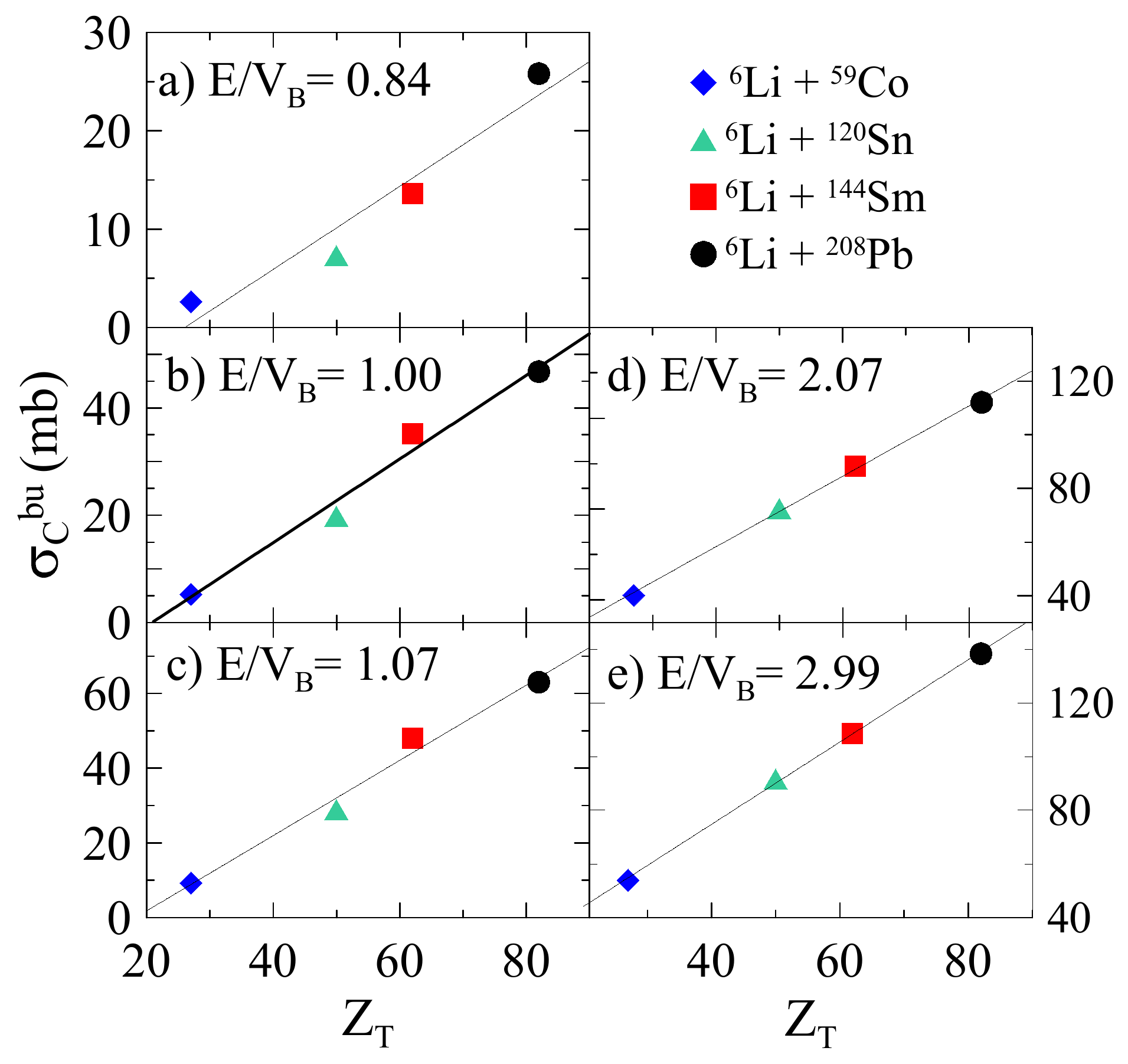}
\end{center}
\caption{(colour on line) Coulomb breakup cross sections plotted as
functions to the atomic numbers of the targets. The notation is the same as
in the previous figure.}
\label{setup}
\end{figure}

\bigskip

Now we consider Coulomb breakup. Fig. 2 shows CDCC calculations of $\sigma_{%
\mathrm{\scriptscriptstyle  C}}^{\mathrm{bu}}$ for the same systems and
collision energies of the previous figure. Now the results are plotted
against the atomic number of the target. We observe that the points are very
well described by the linear fits represented by solid lines. This scaling
for $\sigma _{\mathrm{\scriptscriptstyle  C}}^{\mathrm{bu}}$ can be
understood from the low energy behavior of the Coulomb dissociation cross
section \cite{hussein91}. The electromagnetic coupling matrix-elements are
proportional to $Z_{\mathrm{\scriptscriptstyle  T}}$, which should lead to a 
$Z_{\mathrm{\scriptscriptstyle  T}}^{2}$ dependence in the Coulomb breakup
cross section, whereas the cross sections for reaction channels are
proportional to a $1/E$ factor~\cite{wongnew}. Since in each panel the
collision energy corresponds to the same $E/V_{\mathrm{\scriptscriptstyle  B}%
}$ ratio, and $V_{\mathrm{\scriptscriptstyle  B}}$ is roughly proportional
to $Z_{\mathrm{\scriptscriptstyle  T}}$, one gets a $1/Z_{\mathrm{%
\scriptscriptstyle  T}}$ factor. The combination of these two dependences
leads to the linear dependence obtained in Fig. 2.

\bigskip The above discussion about the scaling of the nuclear and Coulomb
breakup cross sections at near-barrier energies should be useful in the
study of low energy fusion of weakly bound nuclei as it provides means to
assess the feasibility of performing a given experiment which aims to
discern the influence of breakup on fusion.\newline

\noindent

\textbf{Acknowledgements}

\medskip \noindent The authors would like to thank the financial support
from CNPq, CAPES, FAPERJ, FAPESP and the PRONEX.

\end{document}